\documentclass[11pt, oneside]{article}

\usepackage[paper=a4paper, total={16cm,23cm}]{geometry}
\usepackage[T1]{fontenc}
\usepackage{setspace} 
\usepackage[hang,stable]{footmisc} 
\usepackage{titling} 
\usepackage{tocloft} 
\usepackage{parskip} 
\usepackage{comment}
\usepackage{manfnt}
\usepackage{fontawesome}

\usepackage{multirow}

\usepackage{amsfonts}
\usepackage{amsmath}
\usepackage{amssymb}
\usepackage{mathrsfs, dsfont, amsthm}
\usepackage{physics}
\usepackage{mathtools}
\usepackage{upgreek}
\numberwithin{equation}{section} 
\usepackage{wasysym}
\usepackage{marvosym}
\usepackage{cancel}

\usepackage{xcolor}
\usepackage{colortbl}
\definecolor{dark-red}{rgb}{0.50,0.12,0.12} 
\definecolor{mblue}{rgb}{0.30, 0.45, 0.70}
\definecolor{mred}{rgb}{0.70, 0.20, 0.20}
\definecolor{mgray}{rgb}{0.63, 0.63, 0.63}

\usepackage{hyperref} 
\hypersetup{colorlinks=true, allcolors=dark-red, linktoc=all}

\usepackage{cite}

\usepackage{graphicx}
\usepackage{tikz}
\usepackage{tikzlings}
\usetikzlibrary{calc, patterns, decorations, decorations.markings, shapes, positioning, intersections, quotes, angles}
\usepackage[subrefformat=parens]{subcaption}
\usepackage{pgfplots}
\pgfplotsset{compat=newest}
\newcommand{\mathdefault}[1][]{}

\def \rF {\mathrm{F}}

\def \rmc {\mathrm{c}}

\def \bmu {M}
\def \bnu {N}
\def \brho {K}
\def \bsigma {L}

\newcommand{\beq}{\begin{equation}}
\newcommand{\eeq}{\end{equation}}

\DeclareMathOperator{\arctanh}{arctanh}

\newcommand{\ep}{\mathrm{e}}

\newcommand{\diff}{\mathrm{d}}

\newcommand{\fdiff}{\updelta}
\newcommand{\AdS}{\text{AdS}}
\newcommand{\rh}{r_\mathrm{h}}
\newcommand{\zh}{z_\mathrm{h}}

\newcommand{\defeq}{\mathrel{\rlap{\raisebox{0.3ex}{$\cdot$}}\raisebox{-0.3ex}{$\cdot$}}=}

\newcommand{\propsiminn}[2]{\mathrel{\vcenter{
  \offinterlineskip\halign{\hfil$##$\cr
    #1\sim\cr\noalign{\kern0pt}#1\propto\cr\noalign{\kern0pt}}}}}

\begin{document}
\begin{titlingpage}
    \vspace*{3em}
    \onehalfspacing
    \begin{center}
    {\LARGE Boundary imprint of bulk causality} 
    \end{center}
    \singlespacing
    \vspace*{2em}
      \begin{center}
        \textbf{
        Simon Caron-Huot, Joydeep Chakravarty,
        and Keivan Namjou
        }
    \end{center}
    \vspace*{1em}
    \begin{center}
        \textsl{
        Department of Physics, McGill University \\
        Montr\'eal, QC, Canada \\[\baselineskip]
        }
        \href{mailto:schuot@physics.mcgill.ca}{\small schuot@physics.mcgill.ca},
        \href{mailto:joydeep.chakravarty@mail.mcgill.ca}{\small joydeep.chakravarty@mail.mcgill.ca},
        \href{mailto:keivan.namjou@mail.mcgill.ca}{\small keivan.namjou@mail.mcgill.ca}
    \end{center}
    \vspace*{3em}
    \begin{abstract}    
        Motivated by the holographic correspondence, we study the boundary imprint of bulk lightcones in spacetimes with boundaries. These lightcones can be observed whenever a localized event takes place in the bulk. The associated boundary surfaces (hyperboloids) reveal the bulk conformal metric.  We work out a Hamilton-Jacobi description of these surfaces and analyze them in explicit examples. Bulk causality translates into a boundary inclusion property from which the bulk geodesic equation can be derived under some assumptions.
    \end{abstract}
\end{titlingpage}
\tableofcontents

\section{Introduction}
Physics in anti-de Sitter spacetime has a dual description in terms of a boundary conformal field theory \cite{Maldacena:1997re}, which enables one to calculate correlators in certain strongly coupled theories by studying a simpler bulk theory. Some qualitative predictions are particularly striking in real time. Generally, boundary correlators encounter singularities as two points become lightlike separated. Holographic correlators display additional features associated with bulk lightcones and scattering processes happening near a bulk point \cite{Polchinski:1999ry, Giddings:1999jq, Gary:2009ae, Heemskerk:2009pn, Maldacena:2015iua, Komatsu:2020sag}. Singularities along these bulk lightcones do not admit clear explanations in terms of boundary kinematics and consequently serve as a clear test of whether a strongly coupled CFT is holographic or not.

\begin{figure}[t!]
    \centering
    \begin{subfigure}{0.45\textwidth}
    \centering
    \begin{tikzpicture}[scale=.8]
        \filldraw[mblue!50] (0.38,-0.6) -- (.98,1.05) -- (.9,1.06) -- (.3,-.6);
        \filldraw[mblue!50] (0.38,-0.6) -- (-0.82,.88) -- (-0.9,.9) -- (.3,-.6);
        \filldraw[mblue!50] (0.38,-0.6) -- (.67,-1.72) -- (0.59,-1.73) -- (.3,-.6);
        \filldraw[mblue!50] (0.38,-0.6) -- (-.6,-2.36) -- (-.68,-2.37) -- (.3,-.6);
        \draw (-2,2.25) -- (-2.0,-3.3)  (2,-3.3) -- (2,2.25);
        \draw[dashed] (2,-3.3) arc (0:180:2 and 0.25);
        \draw (2,-3.3) arc (0:-180:2 and 0.25);
        \draw (2, 2.25) arc (0:360:2 and 0.25);
        \draw[thick] (-1.99,1.2) node[anchor=east]{\footnotesize $H^+(X)$} .. controls (-1.6,0.9) and (1.6,0.5) .. (1.99,0.7);
        \draw[thick,dashed] (-1.99,1.2) .. controls (-1.6,1.5) and (1.6,1.1) .. (1.99,0.7);
        \draw[thick,dashed] (-1.99,-2.3) node[anchor=east]{\footnotesize $H^-(X)$} .. controls (-1.6,-2.6) and (1.6,-2.2) .. (1.99,-1.8);
        \draw[thick] (-1.99,-2.3) .. controls (-1.6,-1.9) and (1.6,-1.5) .. (1.99,-1.8);
        \node[right] at (0.34,-0.6) {$X$};
        \filldraw[mblue] (0.34,-0.6) circle (2pt);
    \end{tikzpicture}\caption{}\label{fig:4pt-lightraysA}
    \end{subfigure}
    \begin{subfigure}{0.45\textwidth}
    \centering
    \begin{tikzpicture}[scale=0.7]
        \fill[mblue!50] (-2.5,-2.73) -- (-3.08,-0.55) -- (-2.92,-0.52) -- (-2.4,-2.69) -- cycle;
        \fill[mblue!50] (-.4,2.14) -- (-3.07,-0.45) -- (-2.92,-0.52) -- (-.2,2.13) -- cycle;
        \draw[color=mblue, thick] (-1.35,-2.31) -- (-2.93,-0.47) arc (30:210:.075) -- (-1.5,-2.35);
        \draw[thick] (-3,3) to[out=-30,in=-150] (3,3) node[anchor=north west]{\footnotesize $H^+(X)$};
        \draw[thick] (-3,-3) to[out=30,in=150] (3,-3) node[anchor=south west]{\footnotesize $H^-(X)$};
        \draw (-3,-0.5) node[anchor=east]{$X$};
        \filldraw[mblue] (-3,-0.5) circle (2pt);
        \draw (-.3,2.13) node[anchor=south]{\footnotesize $\gamma$}
                (-1.65,-2.35) node[anchor=north]{\footnotesize $P$}
                (-1.02,-2.13) node[anchor=north]{\footnotesize $P^\dagger$}
                (-2.5,-2.73) node[anchor=north]{\footnotesize $\gamma^\dagger$};
    \end{tikzpicture}\caption{}\label{fig:4pt-lightraysB}
    \end{subfigure}
    \caption{(a) The past and future lightcone of a bulk point $X$ define boundary hyperboloids $H^\pm(X)$ along which correlators can display singular features.
    (b) The inclusive expectation value $\expval{P \gamma\gamma^\dagger P^\dagger}{0}$, where $P^\dagger/P$ and $\gamma^\dagger/\gamma$ create/absorb a probe and a pulse of light, respectively. Varying one point holding the other three fixed gives a simple way of observing $H^\pm(X)$.}\label{fig:4pt-lightrays}
\end{figure}
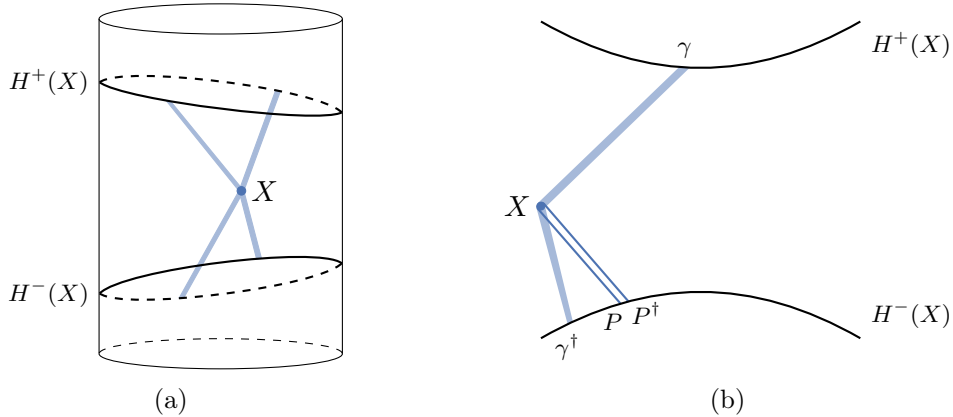

It has been proposed to use the shape of these bulk lightcones to map out the geometry of nontrivial bulk spacetimes. Indeed, the bulk lightcones originate from the rules of geometric optics controlling the propagation of high-frequency waves: why not use optics to decode the hologram? Notably, a precise experimental method to measure the bulk conformal metric has been described \cite{Engelhardt:2016wgb, Engelhardt:2016crc}, assuming one had access to a holographic CFT$_d$ on which one could measure correlation functions. The essential data is distilled in the \emph{boundary hyperboloids} $H^\pm(X)$, which represent the intersection of the future and past lightcone of a bulk point $X$ with the boundary, see figure~\ref{fig:4pt-lightraysA}.\footnote{Our hyperboloids $H^\pm(X)$ are closely related, but not equivalent, to the light-cone cuts $C^\pm(X)$ of \cite{Engelhardt:2016wgb}, see section~\ref{s:inclusion}.} 

We pause to highlight the distinction between ``measurement'' and ``reconstruction'' in the manner of \cite{Fefferman:1985con, Hamilton:2006az}. If a holographic system could be realized in the laboratory and the curves $H^\pm(X)$ measured, it would be straightforward to correctly interpret them and to extract a geometry, using no prior knowledge of bulk equations of motions. This is of course only a thought experiment at the moment--we do not know a concrete realization of a holographic theory. Yet, we hope that consideration of such experiments can add insight, for example, into the mechanism of holography and the emergence of extra dimensions of space and gravity from field theories.

Applications of these ideas have remained relatively scarce. In our opinion, this may be partly due to the relative complexity of the procedure described in \cite{Engelhardt:2016crc}, which required studying correlators of (at least) $d+2$ points \cite{Maldacena:2015iua, Engelhardt:2016wgb}. Intuitively, this is because a local operator inserted at a boundary point emits a non-directional wavefront that spreads into the bulk at the speed of light (considering massless fields). The intersection of $d+1$ such wavefronts generically singles out a unique bulk point, but an additional one must be fine-tuned in order to observe a sharp feature.

More straightforward methods to measure bulk lightcones and their associated boundary hyperboloids have been described recently \cite{Caron-Huot:2022lff}. The \emph{radar camera} described there consists of sending a probe particle with a directional wavepacket that localizes it to a single null geodesic, together with a pulse of light; one then observes the reflected light. The probe and pulse generically intersect at a single bulk point $X$, and the reflected light creates a boundary signal on $H^+(X)$. In this experiment, the final state of the probe is traced over: what happens to the probe after light is reflected off it is not observed.

The essential novel ingredients in this setup are the use of directional wavepackets (also used in early discussions of bulk-point singularities \cite{Polchinski:1999ry, Giddings:1999jq}), and of in-in or ``inclusive'' (non-time-ordered) correlators to trace over the final state of a probe. The radar camera can be modeled as a four-point correlator with three points approaching $H^-(X)$ and one point approaching $H^+(X)$, see figure~\ref{fig:4pt-lightraysB}. While \cite{Caron-Huot:2022lff} focused on certain coincidence limits, a general factorization formula expressing singular features when multiple operators approach the hyperboloid of a common bulk point will be described in coming work \cite{Caron-Huot:2025tap}.

In this paper, we take it as a given that the hyperboloids $H^\pm(X)$ are observable in the boundary theory,  and our goal will be to characterize their geometry. We will work out a general Hamilton-Jacobi description of these surfaces, analyze them in explicit examples, and study how bulk causality constrains their shapes.

Although we focus on asymptotic AdS boundaries, our considerations could apply to other boundaries, real or artificial, such as null infinity in Minkowski space or small tubes surrounding local observers.

\paragraph{Notation and conventions} Throughout, we use uppercase letters for bulk quantities, such as position $X$ and momentum $P$, and lowercase letters for boundary quantities, such as position $x$ and momentum $p$. An exception occurs when the bulk possesses Killing symmetries; in such cases, the bulk and boundary momenta correspond to the same conserved charges, and we use lowercase letters for both.

For indices, we adopt the following conventions: uppercase Latin indices ($\bmu, \bnu, \dots$) are used for $(d+1)$-dimensional bulk (co)vectors; lowercase Greek indices ($\mu, \nu, \dots$) are used for $d$-dimensional quantities, either (co)vectors in the boundary or non-radial components in the bulk; and lowercase Latin indices ($a, b$) are used for spatial components of (co)vectors on the boundary and non-radial spatial components in the bulk. The AdS radius is systematically set to unity: $\ell_\AdS=1$.

\section{Hamilton-Jacobi formalism and null geodesics}\label{s:hj-wkb}
Bulk lightcones can be conveniently characterized by a worldline action that describes a single particle following a geodesic trajectory. The on-shell action, as defined by the Hamilton-Jacobi formalism, provides the leading WKB approximation to the particle's wavefunction inside the bulk.

\subsection{The bulk-boundary Hamilton-Jacobi function}
The dynamics of a relativistic particle of mass $m$ is governed by the worldline action:
\begin{equation}\label{eq:worldlineaction}
    S[X] = - m \int \diff \sigma \, \sqrt{- g_{\bmu\bnu} \frac{\diff X^\bmu}{\diff \sigma} \frac{\diff X^\bnu}{\diff \sigma}}.
\end{equation}
However, this action does not describe massless particles and the square root can be cumbersome for some purposes, such as quantization. This motivates introducing the einbein $\eta$ and rewriting the action as:
\begin{equation}\label{eq:ein-action}
    S[X,\eta] = \frac{1}{2} \int \diff \sigma \, \qty( \frac{g_{\bmu\bnu}}{\eta} \frac{\diff X^\bmu}{\diff \sigma} \frac{\diff X^\bnu}{\diff \sigma} - \eta m^2 ).
\end{equation}
Reparametrization invariance persists in the form $\eta \, \diff \sigma = \eta' \, \diff \sigma'$.

The original worldline action (\ref{eq:worldlineaction}) is recovered by solving for $\eta$ using its equation of motion:
\begin{equation}
    g_{\bmu\bnu} \frac{\diff X^\bmu}{\diff \sigma} \frac{\diff X^\bnu}{\diff \sigma} = - \eta^2 m^2,
\end{equation}
and substituting this solution back into the action (\ref{eq:ein-action}). Let us focus on the value of the on-shell action. Varying the action (\ref{eq:ein-action}) with respect to $X^\bmu$ yields:
\begin{equation}
    \fdiff_X S[X,\eta] = \frac{1}{\eta} \qty[g_{\bmu\bnu} \frac{\diff X^\bmu}{\diff \sigma} \fdiff X^\bnu]_{X_i}^{X_f} - \int \diff \sigma \, \frac{g_{\brho\bsigma}}{\eta} \qty(\frac{\diff^2 X^\bsigma}{\diff \sigma^2}+\Gamma^{\bsigma}_{\bmu\bnu} \frac{\diff X^\bmu}{\diff \sigma} \frac{\diff X^\bnu}{\diff \sigma}) \fdiff X^\brho,
\end{equation}
where $\Gamma^{\bsigma}_{\bmu\bnu}$ are the Christoffel symbols. The second term vanishes when the geodesic equation is satisfied, leaving the variation of the on-shell action as:
\begin{equation}\label{eq:wl-action-on-var}
    \fdiff S(X_i; X_f) = P_f \cdot \fdiff X_f - P_i \cdot \fdiff X_i,
\end{equation}
where the momentum is defined as:
\begin{equation}\label{eq:momentum-def}
    P_\bmu = g_{\bmu\bnu} \frac{\diff X^\bnu}{\diff \lambda}, \qquad \text{where} \qquad \diff \lambda = \eta \, \diff \sigma.
\end{equation}
Here, $\lambda$ is an affine parameter along the geodesic. The reparametrization $\diff \sigma = \eta^{-1} \diff \lambda$ ensures that this momentum definition applies consistently to both massive and massless particles. The momentum satisfies the mass-shell condition, derived from the einbein equation of motion:
\begin{equation}\label{eq:mass-shell}
    g^{\bmu\bnu} P_\bmu P_\bnu + m^2 = 0.
\end{equation}

In the Hamilton-Jacobi formalism, the on-shell action $S(X_i; X_f)$ in (\ref{eq:wl-action-on-var}) is treated as a function of the two endpoints and is to be evaluated by finding a classical solution connecting them. For a massive particle, such a geodesic generically exists and any variation can be considered. There can be situations where multiple geodesics connect the same endpoints; we omit this from our notations since our considerations apply to each solution separately.

In the massless case, the action (\ref{eq:ein-action}) is identically zero on-shell, and (\ref{eq:wl-action-on-var}) is best viewed as a constraint that any on-shell endpoint variation must satisfy:
\begin{equation}\label{eq:wl-action-on-var-massless}
 0 = P_f \cdot \fdiff X_f - P_i \cdot \fdiff X_i\qquad\mbox{(massless case)}.
\end{equation}
In particular, if one considers a family of null geodesics that emanate from a fixed point $X_i$, then the other endpoint can only move in the directions that are orthogonal to $P_f$, i.e., $P_f \cdot \fdiff X_f = 0$. This property of the moduli space of null geodesics will be important in section~\ref{s:geometry}.

As discussed in the introduction, it is natural to integrate boundary correlators against directional wavepackets. An ideal wavepacket has localization in both position and momentum. For the Hamilton-Jacobi problem, it turns out to be most convenient to perform a canonical transformation and fix the boundary momentum. More precisely, we consider geodesics whose first endpoint lies in the asymptotic boundary of spacetime and we fix the components $p_\mu$ of its momentum that are parallel to the boundary:\footnote{We take the boundary to be flat Minkowski space. If the boundary is a nontrivial manifold, $x^\mu$ should be viewed as local coordinates in a neighborhood of the classical path under consideration.}
\begin{equation} \label{SXp}
    S(p; X) \defeq \qty[S(x; X) + p_\mu x^\mu]_{x^\mu = x^\mu(p; X)}
\end{equation}
where $x^\mu(p; X)$ is found by finding a bulk geodesic (if it exists) that passes through the bulk point $X^\bmu$ and reaches the boundary with momentum $p_\mu$.

The action $S(p; X)$ is no longer identically zero for massless particles, and its variations with respect to $X$ are now unconstrained even in the massless case. The boundary momentum does not satisfy any mass shell constraint, but it should normally be timelike.

The complete information about boundary hyperboloids is contained in the Hamilton-Jacobi function (or WKB phase) $S(p; X)$ defined in (\ref{eq:wl-action-on-var-massless}). It depends on a boundary shooting momentum $p_\mu$ and bulk point $X^\bmu$. The canonically conjugate boundary position and bulk momentum are encoded in the derivatives of $S$:
\begin{equation} \label{defs}
    x^\mu(p;X) = \frac{\partial S(p; X)}{\partial p_{\mu}}, \qquad \text{and} \qquad P_M(p;X) = \frac{\partial S(p;X)}{\partial X^{M}}.
\end{equation}
In practice, we will distinguish the past and future branches $H^\pm(X)$ from the sign of $p^0$.

\subsubsection{Geodesic equation from Hamilton-Jacobi\footnote{The discussion in this subsection can be safely skipped on the first reading.}}
As mentioned, the function $S(p;X)$ generally satisfies
\begin{equation} \label{HJ}
    g^{\bmu\bnu}(X)\frac{\partial S}{\partial X^\bmu}\frac{\partial S}{\partial X^\bnu}+m^2=0,
\end{equation}
where $g^{\bmu\bnu}(X)$ is some non-degenerate inverse metric in the bulk. (In most of the paper we set $m = 0$ but we keep it here for future reference.) Here we explain how this relation suffices to recover the bulk geodesic equation.

Starting from the boundary curves $x^\mu(p; X)$, it is natural to look for bulk point displacements $\fdiff X^\bmu$ that leave the boundary endpoint $x^\mu$ and $p_\mu$ unchanged. From the first of (\ref{defs}), this implies that
\begin{equation}
    \frac{\partial^2 S(p;X)}{\partial p_\mu \partial X^\bmu}\fdiff X^\bmu=0\qquad \mbox{for $\fdiff X$ that fixes $x$ and $p$}.
\end{equation}
Now, differentiate (\ref{HJ}) with respect to $p_\mu$:
\begin{equation} \label{dX prop P}
    0=\frac{\partial^2 S(p;X)}{\partial p_\mu \partial X^\bmu} g^{\bmu\bnu}(X)P_\bnu=0.
\end{equation}
Provided that the second derivative has the maximal possible rank, $d$, the two equations imply that
\begin{equation}
    \fdiff X^\bmu \propto g^{\bmu\bnu}(X)P_\bnu.
\end{equation}
Thus, bulk points associated with the same boundary endpoint necessarily move along $P^M$. We would like to show that they follow geodesics. For a given $x^\mu$ and $p_\mu$, consider the one-parameter family $X^\bmu(\lambda)$ of bulk points determined by:
\begin{equation}
    \frac{\diff X^\bmu}{\diff \lambda} = g^{\bmu\bnu}(X) \frac{\partial S(p;X)}{\partial X^\bnu}.
\end{equation}
Differentiating with respect to $\lambda$ gives
\begin{align} \label{ddot equation}
    \frac{\diff^2 X^\bmu}{\diff \lambda^2} = - g^{\bmu\bnu} \frac{\partial g_{\bnu \bsigma}}{\partial X^\brho} \frac{\diff X^\brho}{\diff \lambda} \frac{\diff X^\bsigma}{\diff \lambda} + g^{MN} \frac{\partial^2S}{\partial X^\bnu \partial X^P} \frac{\diff X^P}{\diff \lambda}.
\end{align}
The second term can be simplified using the derivative of the Hamilton-Jacobi equation~(\ref{HJ}):
\begin{align}
    \frac{\partial^2S}{\partial X^\bnu \partial X^\brho} \frac{\diff X^P}{\diff \lambda} &\equiv \frac{\partial^2 S}{\partial X^\bnu \partial X^\brho}
g^{PQ} \frac{\partial S}{\partial X^Q} = - \frac{1}{2} \frac{\partial g^{\brho \bsigma}}{\partial X^\bnu}   \frac{\partial S}{\partial X^\brho} \frac{\partial S}{\partial X^\bsigma} = \frac{1}{2} \frac{\partial g_{\brho \bsigma}}{\partial X^\bnu} \frac{\diff X^\brho}{\diff \lambda} \frac{\diff X^\bsigma}{\diff \lambda}.
\end{align}
Thus (\ref{ddot equation}) is equivalent to the geodesic equation:
\begin{equation}
    \frac{\diff^2 X^\bmu}{\diff \lambda^2} + \Gamma^\bmu_{\brho \bsigma} \frac{\diff X^\brho}{\diff \lambda} \frac{\diff X^\bsigma}{\diff \lambda} = 0.
\end{equation}
We conclude that bulk trajectories $X(\lambda)$ that correspond to a constant boundary endpoint $(x^\mu, p_\mu)$ are precisely the geodesics of the metric $g$ entering the Hamilton-Jacobi equation~(\ref{HJ}).

In the massless case, (\ref{HJ}) only defines the metric up to a local Weyl rescaling---the conformal metric. Since a Weyl rescaling combined with a reparametrization of $\lambda$ is then a symmetry of the geodesic equation, the trajectories themselves are invariant.

\subsection{Bulk depth from boundary hyperboloids}\label{ssec:depth}
\begin{figure}
    \centering
    \begin{tikzpicture}[scale=1.5]
    \draw[fill=mblue!10,opacity=0.3] (0.5,-0.5) -- (3,-1) -- (3,4) -- (0.5,4.5) -- cycle;
    \draw[thick] (0.5,4) .. controls (1,1.3) and (2,1) .. (3,3.5){};
    \draw[color=mblue,decoration={markings,mark=at position 0.45 with \arrow{>}},postaction=decorate, thick] (-.5,0) -- (1.2,2.05);
    \draw[<->] (-.3,.1) arc (30:60:0.3) node[midway,anchor=south east,font=\footnotesize] {$\fdiff P$};
    \draw[color=mblue,dashed] (-.5,0) -- (1.45,1.85);
    \draw[color=mgray,dashed] (.9,1.186) -- (1.45,1.85);
    \draw[<->] (1.25,1.75) arc (-150:-120:0.3) node[midway,anchor=north west,font=\footnotesize] {$\fdiff p$};
    \draw[<->] (1.27,2.1) -- (1.47,1.93) node[midway,anchor=south west,font=\footnotesize] {$\fdiff x$};
    \filldraw[color=mblue] (-.5,0) circle (1pt) node[anchor=north east,font=\footnotesize,color=black]{$X$};
    \filldraw[color=mblue] (1.2,2.05) circle (1pt) node[anchor=east,font=\footnotesize,color=black]{$x$};
\end{tikzpicture}
\caption{Future branch of a boundary hyperboloid $x^\mu(p;X)$. Points along the hyperboloid are labeled by the boundary shooting momentum $p_\mu$, which coincides with the surface normal and determines the bulk momentum $P_\bmu$. Varying $X$ creates a different hyperboloid.}\label{fig:hyperboloid-variation}
\end{figure}
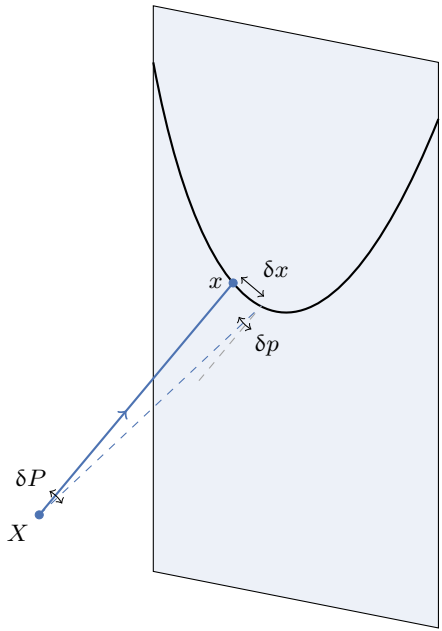

Let us briefly
analyze the geometry of a single boundary hyperboloid $H^\pm(X)$, as determined by the Hamilton-Jacobi function in $S(p;X)$ for fixed $X$, namely (\ref{defs}):
\begin{equation}
    x^\mu(p;X) = \frac{\partial S(p; X)}{\partial p_{\mu}}.
\end{equation}
Since null geodesics are invariant under rescaling of $p$, we have the following two identities:
\begin{equation} \label{hyperboloid normal}
    p_\mu \frac{\partial x^\nu}{\partial p_\mu} = 0 = p_\mu \frac{\partial x^\mu}{\partial p_\nu}.
\end{equation}
The second identity follows from the first by commutativity of the derivatives of $S$ and means that $p$ is orthogonal to all tangent vectors at $x$, as was also noted below (\ref{eq:wl-action-on-var-massless}). Thus, for fixed $X$, the function $x^\mu(p; X)$ parametrizes a codimension-one boundary hypersurface in terms of its normal covector. This is depicted in figure~\ref{fig:hyperboloid-variation}. (There could be isolated points where this parametrization is singular but we assume that it generically applies locally.)

Even though null geodesics only probe the bulk metric up to Weyl rescaling, boundary distances offer various notions of the ``depth'' of the bulk point. A simple one is the back-and-forth travel time: the time separation between the past and future hyperboloids from $X$. One could also explore more covariant notions, such as the minimal proper time between one point in the past hyperboloid of $X$ and one in its future, although it is not clear that these are always useful.

A second simple notion of depth, which underlies stereographic vision, is the parallax $\partial x^\mu / \partial p_\nu$: how the arrival position of light from $X$ correlates with its arrival angle. The parallax turns out to be closely related to the extrinsic curvature of the boundary hyperboloid. The general formula for the extrinsic curvature of a surface is:\footnote{Here, we treat the boundary momenta as coordinates on the hyperboloid, and to avoid confusion, we use dotted indices for these coordinates.}
\begin{equation}
    \tilde{K}^{\dot{\mu} \dot{\nu}} (p,X)= -n_\rho \nabla^{\dot{\mu}} \frac{\partial x^\rho}{\partial p_{\dot{\nu}}} = - n_\rho \qty(\frac{\partial^2 x^\rho}{\partial p_{\dot{\mu}} \partial p_{\dot{\nu}}} + \Gamma^\rho_{\mu \nu} \frac{\partial x^\mu}{\partial p_{\dot{\mu}}} \frac{\partial x^\nu}{\partial p_{\dot{\nu}}}),
\end{equation}
where here $n_\rho = p_\rho / \sqrt{-p^2}$ is the unit normal to the curve. The Christoffel symbol $\Gamma^\rho_{\mu\nu}$ vanishes for a flat Minkowski boundary, and by adding a derivative of (\ref{hyperboloid normal}), we can simplify the above expression as
\begin{equation}
    \tilde{K}^{\dot{\mu} \dot{\nu}} = \frac{1}{\sqrt{-p^2}}\frac{\partial x^{\dot{\mu}}(p;X)}{\partial p_{\dot{\nu}}}=
    \frac{1}{\sqrt{-p^2}}\frac{\partial^2 S(p;X)}{\partial p_{\dot{\mu}}\partial p_{\dot{\nu}}}\qquad\mbox{(Minkowski boundary)}.
\end{equation}
In other words, the apparent depth of a bulk point (as would be perceived from two adjacent eyes at the boundary) is the curvature radius of the hyperboloid. It is not necessarily constant nor isotropic along the hyperboloid.

\section{Examples of boundary hyperboloids}\label{s:geometry}
Each bulk point $X$ uniquely determines a set of points on the conformal boundary that are lightlike separated from it, forming what we refer to as boundary hyperboloids $H^\pm(X)$. These hyperboloids have two branches: the future branch, representing the projection of the point's future lightcone onto the conformal boundary, and the past branch, arising from the past lightcone. As we will illustrate in this section and the next one, the collection of these hyperboloids reveals a great deal of information about the conformal geometry and causal structure of the bulk. The metrics we consider in this section can be found in \cite{Ammon:2015wua}.

\subsection{Translational invariant examples}
While the constructions described so far are rather general, here we specialize for illustration to a general class of metrics with translation, spatial rotation, and time-reversal isometries:
\begin{equation}\label{eq:pbhmetric}
    \diff s^2 = \frac{1}{z^2} \qty(-A(z) \, \diff T^2 + \delta_{ab} \, \diff X^a \diff X^b + \frac{\diff z^2}{B(z)}).
\end{equation}
The boundary of the spacetime is at $z = 0$, where $A=B=1$; unless specified otherwise, we set the AdS radius $\ell_\AdS = 1$ throughout. This class of metrics includes black brane solutions, for example.

Motion in this geometry is governed by the translation Killing vectors, which align with the non-radial directions $\partial_{\mu}$ and ensure that the momenta $p_\mu=g_{\mu\bmu}\frac{d X^\bmu}{d\lambda}$ are constant along geodesics. For a massless particle moving toward the boundary, the mass-shell condition in (\ref{eq:mass-shell}) then gives:
\begin{equation} \label{mass-shell-modified}
    P_z = -\sqrt{-g_{zz}(z) g^{\mu\nu}(z) p_\mu p_\nu}
\end{equation}
which is generally a nontrivial function of $z$. Choosing the radial coordinate as the integration variable, the Hamilton-Jacobi function and boundary coordinates are then
\begin{equation} \label{eq:geo-int}
    S(p;X) = p_\mu X^\mu - \int_0^z \diff z\,\sqrt{-g_{zz} g^{\mu\nu} p_\mu p_\nu}, \qquad x^\mu(p;X) = X^\mu + \int_0^z \diff z \, \frac{g^{\mu\nu} p_\nu}{\sqrt{-g^{zz} g^{\rho\sigma} p_\rho p_\sigma}}.
\end{equation}
The function $x^\mu(p; X)$ specifies the boundary hyperboloid from the point $X$ in terms of the boundary momentum $p$. The future and past branches of this hyperboloid are distinguished here by the sign of the temporal component $E$ of the momentum $p^\mu=(E,\mathbf{p})$.

In the following, we discuss explicit examples of translation-invariant geometries to develop intuition for the hyperboloids.

\subsubsection{Poincar\'e AdS}
We recover the case of empty AdS in Poincar\'{e} coordinates by setting $A(z) = B(z) = 1$ in the metric (\ref{eq:pbhmetric}):
\begin{equation}\label{eq:poincaremetricf}
    \diff s^2 = \frac{1}{z^2} \qty(-\diff T^2 + \delta_{ab} \, \diff X^a \diff X^b + \diff z^2).
\end{equation}
 In these coordinates,
the integrands in (\ref{eq:geo-int}) are independent of $z$ and the boundary hyperboloid is characterized by:
\begin{equation}
    t(p;X) = T + \frac{E z}{\sqrt{E^2 - \mathbf{p}^2}},\qquad
    x^a(p;X) = X^a + \frac{p^a z}{\sqrt{E^2 - \mathbf{p}^2}}.
\end{equation}
This geometry is dual to the CFT vacuum and enjoys more symmetries compared to an arbitrary state, a property that is partly reflected in the geodesics. Eliminating the parametric dependence on boundary momenta $p$, the curve is simply a hyperbola:
\begin{equation} \label{hyperbola}
    - (t - T)^2 + \delta_{ab} (x^a - X^a)(x^b - X^b) + z^2 = 0.
\end{equation}
This formula should not be surprising since we are studying null geodesics and the metric (\ref{eq:poincaremetricf}) is conformally flat. We see that the curvature radius of the hyperbola is precisely the radial depth $z$ of the bulk point in Poincar\'e coordinates, as anticipated qualitatively in \S\ref{ssec:depth}.

\subsubsection{Planar black holes} 
Let us now consider the boundary hyperboloids causal to a bulk point in a black brane geometry. The metric components of a planar black hole (back brane) in asymptotically AdS space take the form (\ref{eq:pbhmetric}) with:
\begin{equation}
    A(z) = B(z) = 1 - \frac{z^d}{\zh^d}.
\end{equation}
In these coordinates, the black hole horizon is at $z = \zh$, while $z< \zh$ covers the black hole exterior and the conformal boundary lies at $z = 0$. The Hawking temperature of this black hole is given by $\beta = {4 \pi z_h / d}$. The geodesics that reach the boundary must originate from a bulk point $X$ in the exterior region ($z < z_h$) and move radially outward towards $z=0$. Note that empty AdS in Poincar\'{e} coordinates can be viewed as the small-radius limit $\zh \to \infty$ of this geometry.

We first study the hyperboloids for $d=2$, i.e., a $2+1$ dimensional bulk corresponding to the BTZ black hole, where the components in the metric (\ref{eq:pbhmetric}) are $A(z) = B(z) = 1 - z^2/\zh^2$ and the integrals (\ref{eq:geo-int}) give:
\begin{equation}
\begin{split}
    t(p;X) &= T + \zh \arctanh\qty(\frac{z}{\zh} \frac{E}{\sqrt{E^2 - p^2 A(z)}}), \\
    x(p;X) &= X + \zh \arctanh\qty(\frac{z}{\zh} \frac{p}{\sqrt{E^2 - p^2 A(z)}}).
\end{split}
\end{equation}
However, these may not be the most exciting formulas to plot since the $d=2$ geometry is locally equivalent to empty AdS: these curves are ultimately just a conformal transformation of a hyperbola.

\begin{figure}
    \centering
    \input{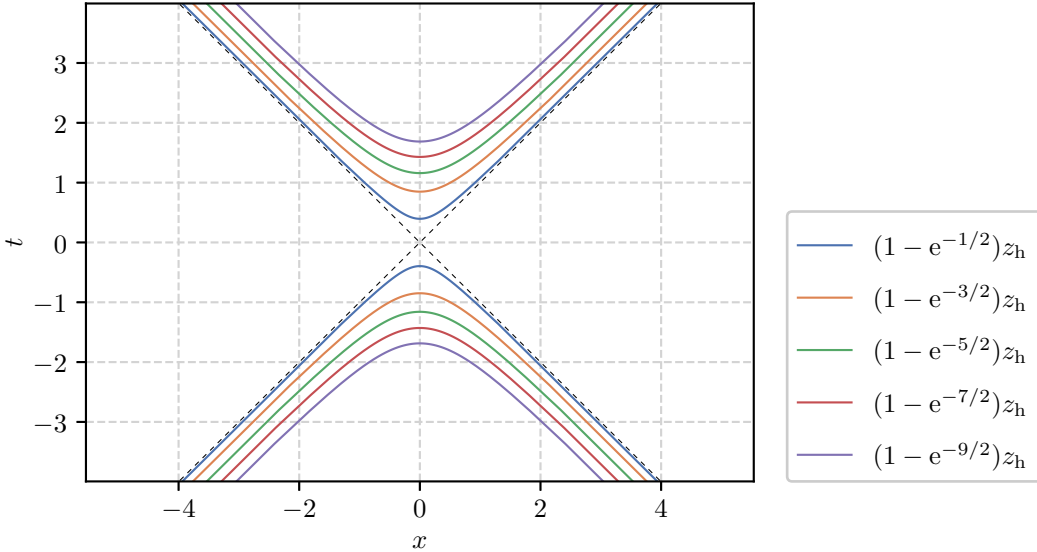}
    \caption{Boundary hyperboloids corresponding to various shooting radii $z = (1 - \ep^{-n/2}) \zh$ outside a planar AdS$_5$ black hole with horizon radius $\zh = 1$. The curves become actual hyperbolas as $z\to 0$ but become time-translates of a common shape as $z\to \zh$.}\label{fig:planar}
\end{figure}

We thus 
consider the AdS$_5$ planar black hole, for which we have $A(z) = B(z) = 1 - z^4/\zh^4$ (the exercise could easily be repeated in any other dimension). Using these components, we perform the integrals (\ref{eq:geo-int}) and obtain the following parametric form for the boundary hyperboloids parametrized by boundary momenta:
\begin{equation}
\begin{split}
    t(p;X) &= T + \frac{E z}{\sqrt{E^2 - \mathbf{p}^2}} \, \rF_1\qty(\frac{1}{4}, \frac{1}{2}, 1, \frac{5}{4}; - \frac{\mathbf{p}^2}{E^2 - \mathbf{p}^2} \frac{z^4}{\zh^4}, \frac{z^4}{\zh^4}), \\
    x^a(p;X) &= X^a + \frac{p^a z}{\sqrt{E^2 - \mathbf{p}^2}} \, {}_2\rF_1\qty(\frac{1}{4}, \frac{1}{2}, \frac{5}{4}; - \frac{\mathbf{p}^2}{E^2 - \mathbf{p}^2} \frac{z^4}{\zh^4}),
\end{split}
\end{equation}
where $\rF_1$ and ${}_2\rF_1$ are Appell and hypergeometric functions, respectively. Figure~\ref{fig:planar} presents the corresponding hyperboloids for points at various depths outside the horizon of a planar AdS$_5$ black hole.

As the bulk depth $z$ increases, one can see that the curvature radius of the hyperboloid initially increases until it saturates: the parallax depth becomes constant near the horizon, as was also observed in \cite{Caron-Huot:2022lff}. As $z\to z_h$, the curves stop deforming and become approximately time-translates of each other, as $t$ acquires a large and $p$-independent logarithmic shift. The logarithmic growth can be seen most simply in the back-and-forth time to $X$ along a radial geodesic, which is twice the tortoise coordinate $z_*$:
\begin{equation} \label{tortoise}
    z_* = \int_0^z \frac{\diff z}{A(z)} = z \, {}_2\rF_1\left(1,\frac{1}{d};1+\frac{1}{d};\frac{z^d}{z_\mathrm{h}^{d}}\right) \to \frac{z_h}{d}\log \frac{1}{1-z/z_h} \quad \mbox{as} \quad z\to z_h.
\end{equation} 

\subsection{Small AdS Schwarzschild black holes}
Analyzing the hyperboloids corresponding to the Schwarzschild black hole in AdS offers valuable insights, as it introduces a key qualitative difference from planar cases: geodesics can orbit the black hole before eventually reaching the boundary. To explore this behavior, we study geodesics that circle the black hole. Some properties of such geodesics have been previously studied in \cite{Berenstein:2020vlp, Cruz:1994ir, Hashimoto:2018okj, Hashimoto:2019jmw, Kaku:2021xqp, Dodelson:2022eiz, Kinoshita:2023hgc, Dodelson:2023nnr} as they lead to many remarkable features in two-point functions. Here, our main goal is to understand the properties of the boundary hyperboloids corresponding to a bulk point in the presence of spherical black holes.

For display purposes, it will be convenient to work with a radial coordinate $r=1/z$, such that the spherically symmetric AdS-Schwarzschild black hole takes the form:
\begin{equation}\label{eq:schbhmetric}
    \diff s^2 = - \tilde{A}(r) \diff T^2 + r^2 \diff \Omega_{d-1}^2 + \frac{\diff r^2}{\tilde{B}(r)},
\end{equation}
where, for a Schwarzschild black hole in $d + 1$ dimensions,
\begin{equation}
    \tilde{A}(r) = \tilde{B}(r) = 1 + r^2 - (1 + \rh^2) \, \frac{\rh^{d-2}}{r^{d - 2}}.
\end{equation}
Here $\rh$ is the radius of the event horizon and the asymptotic boundary lies at $r\to \infty$.

With no loss of generality, we can assume that the angular part of the motion takes place within an equatorial circle, with angular momentum $J = r^2 \frac{\diff \theta}{\diff \lambda}$. We denote the bulk point by the coordinates $X = (T, \Theta, r)$, while the boundary coordinates are $x = (t, \theta)$.

As in the planar case, the conserved angular momentum allows us to find $S(p; X)$ in terms of an integral representation similar to (\ref{eq:geo-int}), which essentially gives us the geodesics.
The boundary endpoints are then given as 
\begin{equation}
\begin{split}
    \theta(p;X) &= \Theta + \int_{r}^{\infty} \diff r \, \frac{J}{r^2 \sqrt{E^2 - J^2 \tilde{A}(r) / r^2}}, \\
    t(p;X) &= T + \int_{r}^{\infty} \diff r \, \frac{E}{\tilde{A}(r) \sqrt{E^2 - J^2 \tilde{A}(r) / r^2}}.
\end{split}
\end{equation}
We can also change the integration upper limit to some $r'$ in order to study the bulk geodesics themselves.

\begin{figure}
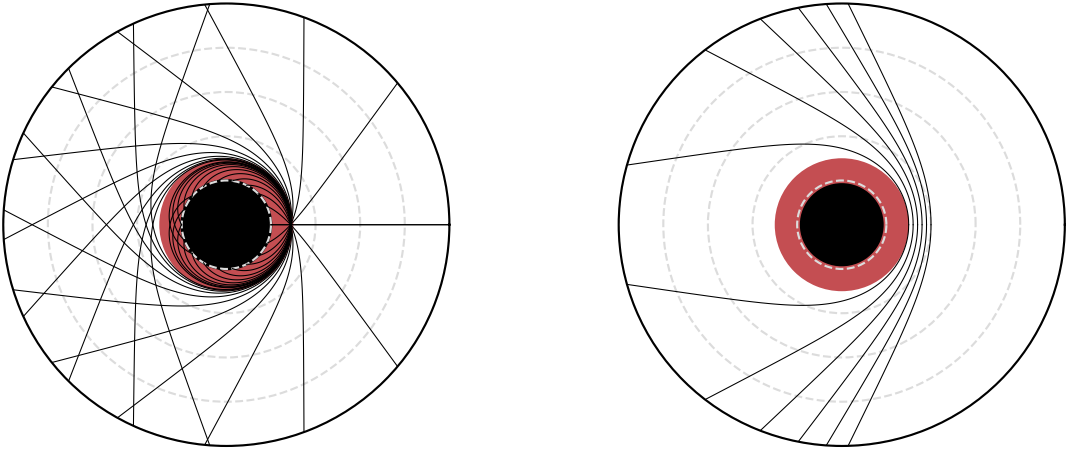

    \centering
    \begin{subfigure}[c]{0.45\textwidth}
        \input{schw-geodesics-bs.pgf}
    \end{subfigure}
    \qquad
    \begin{subfigure}[c]{0.45\textwidth}
        \input{schw-geodesic-turnings.pgf}
    \end{subfigure}
    \caption{Null geodesics in the AdS$_5$-Schwarzschild black hole with $\rh = 0.2$. The black region represents the black hole interior and the red region indicates the interior of the photon sphere at $r_{\mathrm{ps}}$. The left figure shows the geodesics originating from a point at the photon sphere. The figure on the right shows several geodesics that have a turning point due to the presence of the black hole.}
    \label{fig:ads-schw}
\end{figure}

A particularly interesting region is the photon sphere, which is the radius at which the effective potential in the geodesic equation, $V_{\text{eff}} = J^2 \tilde{A}(r) / r^2$, is minimized. There exists a critical value of angular momentum, $J_\rmc$, such that the geodesic equation is well behaved at $r_{\mathrm{ps}}$:
\begin{equation}\label{eq:j-critical}
    \frac{J_\rmc}{E} = \sqrt{\frac{r_\mathrm{ps}^2}{\tilde{A}(r_\mathrm{ps})}}.
\end{equation}
This is the angular momentum of a massless particle which would orbit the black hole indefinitely at the photon sphere (which is of course an unstable orbit).
Specializing to $d=4$ for now on, for example
\begin{equation}
    r_{\mathrm{ps}} = \rh \sqrt{2 \qty(1 + \rh^2)}, \qquad \text{and} \qquad \frac{J_\rmc}{E} = \frac{2 \rh \sqrt{1 + \rh^2}}{1 + 2 \rh^2}.
\end{equation}
Figure~\ref{fig:ads-schw} provides a top-down view of these geodesic trajectories, illustrating the unstable nature of the geodesics near the photon sphere. Trajectories that reach the boundary after a large variation in $\theta$ are possible but they require $J$ to be exponentially fine-tuned near $J_c$.

\begin{figure}
    \centering
    \input{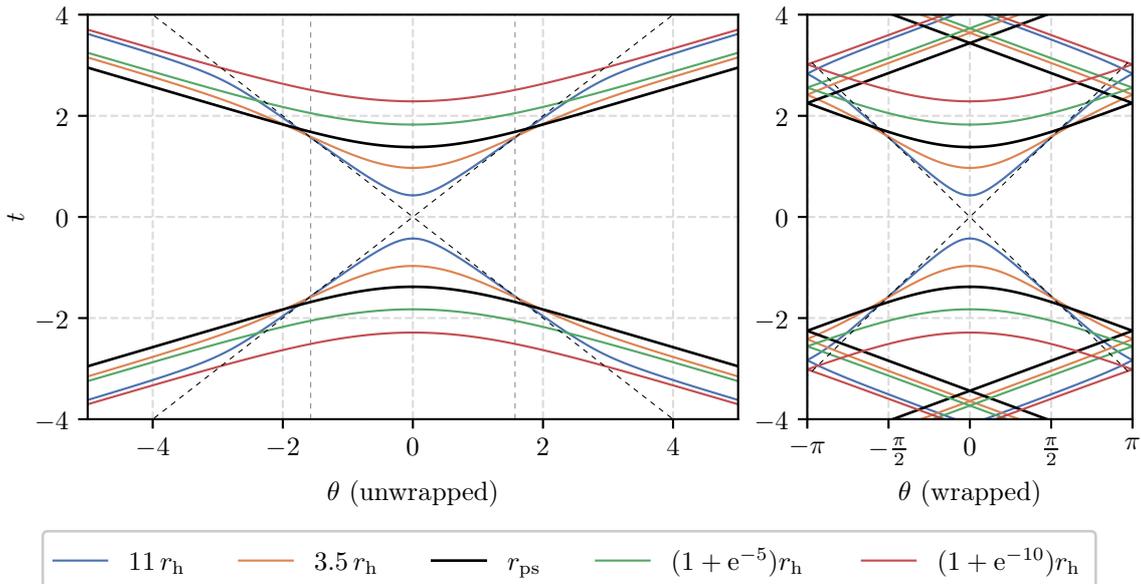}
    \caption{Boundary hyperboloids corresponding to points at different radial coordinates in the presence of a spherical black hole with $\rh = 0.2$. The asymptotic slope of all the hyperboloids is $J_c$. The left plot shows the unwrapped hyperboloid. The right plot illustrates what an observer at the boundary would see, accounting for the geodesics that rotate around the black hole before reaching the boundary.}\label{fig:spherical-hyperboloids-multiple}
\end{figure}

Since we are parametrizing the curves by their normal, the slope of the boundary curves is simply related to $J$ and $E$:
\begin{equation}
    \frac{\diff t}{\diff \theta}\Big|_{\rm boundary} =  \frac{J}{E}.
\end{equation}
The resulting boundary hyperboloids, illustrated in figure~\ref{fig:spherical-hyperboloids-multiple},  resemble those from our planar examples, with two notable distinctions: (1) since $\theta$ is periodic ($\theta \sim \theta + 2 \pi$), the hyperboloids have infinitely many branches corresponding to geodesics that orbit the black hole multiple times before reaching the boundary; (2) the asymptotic value of $J$ is limited by the black hole's properties.

For any radius $r$, there is a maximal angular momentum that a real geodesic can have:
\begin{equation}
    J_{\rm max}(r) = E \sqrt{\frac{r^2}{\tilde{A}(r)}}.
\end{equation}
There is a qualitative difference between bulk points that are inside or outside the photon sphere. For $r>r_{\rm ps}$, one can either shoot outward or inward. Outward geodesics can have any value $0\leq |J| <J_{\rm max}$(r). Inward geodesics with $J_c<|J|<J_{\rm max}(r)$ will reach some minimal radius $r_{\rm min}(J)$ before turning around; as $|J|\to J_c$, these geodesics spend a lot of time orbiting the photon sphere, which leads to a common asymptotic slope $J_c/E$ for all the curves shown in figure~\ref{fig:spherical-hyperboloids-multiple}. The inward and outward cases are connected by an inflection point where the slope $J=J_{\rm max}(r)$ is maximal.

Starting from bulk points that are inside the photon sphere, the only way to reach the boundary is to shoot outward, with angular momentum $0\leq |J|<J_c$; otherwise the geodesic will reflect at some radius $r<r_{\rm ps}$ and fall back into the black hole. This leads to the curves with monotonic slopes visible in figure~\ref{fig:spherical-hyperboloids-multiple}. As $r\to r_h$, the boundary hyperboloids become time-translates of a common curve with a logarithmically increasing time delay, as for the planar black hole discussed previously.

An apparent ``paradox'' is worth mentioning when looking at figure~\ref{fig:spherical-hyperboloids-multiple}: some points on the upper and lower hyperboloids are spacelike separated. This suggests that a bulk path (namely, the union of two null geodesics) could connect these two points faster than any boundary path, in tension with boundary causality and the Gao-Wald theorem \cite{Gao:2000ga}. However, the resolution lies in the fact that the angular coordinate $\theta$ is (necessarily) periodic, as shown in the right panel.\footnote{In $d=2$, corresponding to the BTZ black hole, the angle does not need to be periodic however there is no photon sphere and no analogous paradox.} The most dangerous path connects antipodal points where $\theta$ differs by $\pi$ and so moves in the range $-\pi/2<\theta<\pi/2$. However, in this range, the hyperboloids remain timelike from the origin and the fastest path is along the boundary, as it should.

Finally, we briefly comment on the limiting case of (\ref{eq:schbhmetric}) where we take the horizon radius to be zero and recover the metric of empty global AdS. As opposed to the Poincar\'{e} coordinates which cover only a part of empty AdS, these coordinates cover the entire AdS spacetime. The boundary hyperboloid $H^+$ now has a single ellipsoid branch, which becomes spherical when the bulk point is at the center of AdS (see figure~\ref{fig:4pt-lightraysA}).

\section{Causality and inclusion of hyperboloids}\label{s:inclusion}
So far in this paper, we have studied hyperboloids $H^\pm(X)$ that include all light rays leaving a bulk point $X$ and reaching the boundary. This is the right concept to describe bulk-point-type singularities in boundary correlators. In this section, we will be interested in causal properties of the bulk spacetime, which are best captured by a more refined concept: the light-cone cuts $C^\pm(X)$.

The light-cone cut $C^+(X)$ is the intersection between the boundary of the future of $X$ and the asymptotic boundary;  $C^-(X)$ is the same for the past \cite{Engelhardt:2016wgb}. All points in $C^\pm(X)$ can be reached by a \emph{prompt} null geodesic from $X$, meaning that they cannot also be reached by a piecewise sequence of timelike geodesics \cite{Witten:2019qhl}. Thus, in general, light-cone cuts are a proper subset of our hyperboloids: $C^\pm(X) \subset H^\pm(X)$.

For example, in figure~\ref{fig:spherical-hyperboloids-multiple}, $C^\pm(X)$ only includes the geodesics that reach the bulk in the shortest time for any angle $\theta \in (-\pi,\pi)$, whereas the ``wrapped'' geodesics are not prompt and are only part of $H^\pm(X)$.

A more basic example is a mirror. In the presence of a mirror, we include reflected light rays in $H^\pm$ (they are certainly important for understanding light signals in the presence of the mirror!), but they are not included in $C^\pm$ since they are not prompt nor relevant to the causal structure of the spacetime.

\subsection{Simple method to measure the bulk conformal metric}\label{s:geometry:hyperboloids}
In a causal spacetime, causal relations are transitive: if $Y$ is in the future of $X$ (meaning that they can be connected by a timelike geodesic or sequence thereof), and $Z$ is in the future of $Y$, then $Z$ is automatically in the future of $X$ \cite{Witten:2019qhl}. As was argued in \cite{Engelhardt:2016crc}, this property enables one to extract the bulk conformal metric from a family of lightcone cuts $C^\pm(X)$; here we will spell out an explicit procedure.

We will consider the infinitesimal version of the statement: if $\fdiff X^\bmu$ is a future-directed timelike or null tangent vector at $X$, then the future of $X+\fdiff X$ is contained within that of $X$. In particular, all points in its boundary must move forward:
\begin{equation} \label{inclusion infinitesimal}
     -p_\nu \frac{\partial x^\nu(p;X)}{\partial X^\bmu} \fdiff X^\bmu \geq 0 \quad \forall \, p \qquad \mbox{(timelike or null $\fdiff X$).}
\end{equation}
This equation is remarkable because it means that an infinite number of inequalities (one for each $p$) must be implied by a single inequality on the $(d+1)$ components of $\fdiff X^\bmu$.

To better understand the implications of (\ref{inclusion infinitesimal}), it is useful to ask when it is saturated. Let us first analyze this solely from the boundary perspective. If a variation of a codimension-one surface has a fixed point and satisfies an inequality \eqref{inclusion infinitesimal}, then the displaced and original surfaces must be tangent at this point. Since we are parametrizing the surface by its normal, this means that at a fixed point,
\begin{equation} \label{inclusion saturation}
    \frac{\partial x^\nu(p;X)}{\partial X^\bmu} \fdiff X^\bmu =0
    \qquad\mbox{when (\ref{inclusion infinitesimal}) is saturated}.
\end{equation}
For generic $p$, the $d\times (d+1)$-dimensional matrix $\partial x^\nu/\partial X^\bmu$ has a single null vector (proportional to the bulk momentum $P^\bmu$ of the corresponding geodesic). Thus, from the boundary perspective, for any point $x(p; X)\in C(X)$, one can produce a displacement $\fdiff X^\bmu$ that saturates (\ref{inclusion infinitesimal}) by finding the null space of a matrix.

From the bulk perspective, a vector that saturates (\ref{inclusion infinitesimal}) is necessarily null: if a timelike displacement of a bulk point $X$ was such that $C^+(X)$ and $C^+(X+\fdiff X)$ contained a common point, then that point would not be on a prompt geodesic, which is a contradiction. It is easy to see why a null displacement saturates (\ref{inclusion infinitesimal}): if a null geodesic connects $X$ to $x$, then displacing $X$ along its direction will not affect its endpoint. Thus, combining with (\ref{inclusion saturation}), we obtain:
\begin{equation} \label{saturate}
    \frac{\partial x^\nu(p;X)}{\partial X^\bmu} \fdiff X^\bmu =0\quad\Rightarrow\quad g_{\bmu\bnu}(X)
    \fdiff X^\bmu \fdiff X^\bmu=0.
\end{equation}
This equation can be used to ``measure'' algebraically the bulk metric at a point $X$, up to an overall factor, from small variations of its boundary hyperboloid. The result is vastly over-constrained: in a $(d+1)$-dimensional bulk, $d(d+3)/2$ discrete values of $p$ generically suffice to fully determine $g_{\bmu\bnu}(X)$ up to an overall factor.

For a null displacement that is along a geodesic that is not prompt, ie. a point in $H(X)$, we still expect the equality (\ref{saturate}) to hold, however, the inequality (\ref{inclusion infinitesimal}) may fail to be satisfied for all $p$. 

\begin{figure}
    \centering
    \input{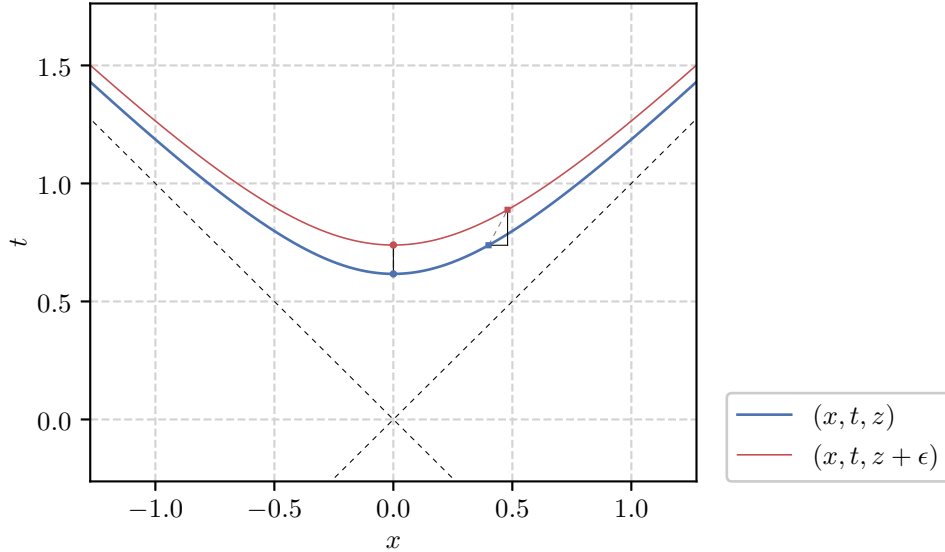}
    \caption{The bulk conformal metric can be read off algebraically by locating points on nearby hyperboloids that have the same normal, see (\ref{black brane reconstruction}). Here we show the boundary hyperboloid corresponding to a point at $z = 0.6$ outside the planar AdS$_5$ black hole with $\zh=1$, and its displacement with $\epsilon = 0.1$.}\label{fig:planar-variation}
\end{figure}

\paragraph{Black brane example}
We illustrate the procedure in an example where translation, time-reversal, and rotational symmetries reduce the number of independent metric components:
\begin{equation}
    \diff s^2 = -g_{TT}(z) \diff T^2 + g_{zz}(z) \diff z^2 + g_{XX}(z) \delta_{ab} \diff X^a \diff X^b\qquad\mbox{(symmetrical example)}.
\end{equation}
Since $\partial x^\nu/\partial X^\mu=\delta^\nu_\mu$, the null vector and constraint associated with a given $p$ are simply
\begin{equation} \label{black brane reconstruction}
   \qty(\frac{\partial t(p;X)}{\partial z},\frac{\partial x(p;X)}{\partial z},-1)\quad\mbox{is null}\quad\Rightarrow\quad -g_{TT}\qty(\frac{\partial t}{\partial z})^2 + g_{XX}\qty(\frac{\partial x}{\partial z})^2+g_{zz}=0.
\end{equation}
Only two evaluations suffice to determine all unknowns, as depicted in figure~\ref{fig:planar-variation}. The graphical procedure involves finding points on the original and displaced curves $C(z)$ and $C(z+\fdiff z)$ whose slopes match. Using just the arrival time at the symmetrical point ($\vec{p}=0$) one finds the ratio $g_{zz}/g_{TT}$, and using any other point one can then determine $g_{XX}/g_{zz}$.  We checked numerically  that this procedure applied to the family of curves in figure~\ref{fig:planar} recovers the conformal metric of the AdS$_5$ planar black hole, namely:
\begin{equation}
    \frac{-g_{TT}}{g_{XX}}= \frac{g_{XX}}{g_{zz}} = 1 - \frac{z^4}{\zh^4}.
\end{equation}
In general, the outcome of this exercise depends on the choice of a radial coordinate $z$ used to parametrize the family of curves. An experimentally natural choice would be to take $z$ equal to half the back-and-forth time to the point $X$, e.g., the tortoise coordinate (\ref{tortoise}). Then $\frac{g_{zz}}{-g_{TT}}=1$ by definition and the outcome of the measurement is the function $\frac{g_{XX}}{g_{zz}}(z_*)$.

\subsection{Reversing the logic: null geodesics in the bulk from causality}\label{ssec:reverse}
It seems that the logic of the preceding subsection can be reversed, allowing one to recover the geometry in the bulk by looking at the boundary hyperboloids.

As a warm-up to the general case, consider a general translation-invariant state and assume that a one-parameter family of surfaces $x^\mu(p;z)$ (representing the boundary hyperboloid of the point $(0,z)$) exists such that, for any $z$, the tangent condition (\ref{inclusion saturation}) is a \emph{quadratic} constraint on bulk displacements. This means that there exists functions $A_{\mu\nu}(z)$ and $B_\mu(z)$ such that: 
\begin{equation} \label{bulk metric assumption}
    \exists \, A_{\mu\nu}(z), B_{\mu}(z), \quad \mbox{such that} \quad A_{\mu\nu} \qty( \frac{\partial x^\mu(p;z)}{\partial z} + B^\mu) \qty( \frac{\partial x^\nu(p;z)}{\partial z} + B^\nu) + 1 = 0, \quad \forall \, p.
\end{equation}
Here we chose a convenient parametrization of the quadratic form (with no loss of generality provided that $A$ is nondegenerate) and we are also assuming (again with no loss of generality) that the surface is parametrized by its normal, so that
\begin{equation} \label{normal condition 2}
    p_\mu \frac{\partial x^\mu(p;z)}{\partial p^\nu} =0.
\end{equation}

We claim that (\ref{bulk metric assumption}) and (\ref{normal condition 2}) determine $\partial x^\mu(p;z)/\partial z$. To see this, take a $p$ derivative of (\ref{bulk metric assumption}):
\begin{equation} \label{bulk metric assumption prime}
    2A_{\mu\nu}(z)\qty( \frac{\partial x^\nu(p;z)}{\partial z} + B^\nu(z)) \frac{\partial^2 x^\mu(p;z)}{\partial p_\sigma \partial z} = 0.
\end{equation}
On the other hand, the $z$ derivative of (\ref{normal condition 2}) gives:
\begin{equation}
    p_\mu \frac{\partial^2 x^\mu(p;z)}{\partial p_\sigma \partial z} = 0.
\end{equation}
For different choices of $\sigma$, this generically gives $(d-1)$ independent constraints. Comparing the preceding two equations we conclude that the parenthesis in (\ref{bulk metric assumption prime}) must be proportional to $p_\mu$, with a proportionality factor that can be determined from (\ref{bulk metric assumption}):
\begin{equation} \label{bulk metric step}
   A_{\mu\nu}(z) \qty(\frac{\partial x^\nu(p;z)}{\partial z} + B^\nu(z)) = \pm \frac{p_\mu}{\sqrt{- (A^{-1}(z))^{\rho\sigma}p_\rho p_\sigma}}.
\end{equation}
Thus from (\ref{bulk metric assumption}) and (\ref{normal condition 2}) one can isolate $\frac{\partial x}{\partial z}$ and derive that 
\begin{equation} \label{bulk metric reconstructed}
    x^\mu(p;z) = \int_0^z \diff z' \, \qty(\frac{\pm (A^{-1}(z'))^{\mu\nu}p_\nu}{\sqrt{- (A^{-1}(z'))^{\rho\sigma}p_\rho p_\sigma}}-B^\mu(z')).
\end{equation}
Thus, the family of surfaces $x^\mu(p;z)$ can be fully recovered from the functions $A$ and $B$ in (\ref{bulk metric assumption})! (In (\ref{bulk metric reconstructed}) we also assumed the natural boundary condition $x^\mu(p;0)=0$ at the AdS boundary.)

It remains to check that this is equivalent to solving the geodesic equation in the conformal metric associated with (\ref{bulk metric assumption}), namely:
\begin{equation} \label{bulk metric ansatz}
    g_{\bmu \bnu} \diff x^\bmu \diff x^\bnu \propto \diff z^2 + A_{\mu\nu}(z) \qty(\diff x^\mu+B^\mu(z)dz) \qty(\diff x^\nu+B^\nu(z)dz).
\end{equation}
Introducing a suitable affine parameter $\diff \lambda = \pm \diff z / \sqrt{-(A^{-1})^{\mu\nu}p_\mu p_\nu}$, (\ref{bulk metric step}) can be rewritten as:
\begin{equation}
    p_\mu = A_{\mu\nu}\qty(\frac{\partial x^\nu}{\partial\lambda} +B^{\nu}\frac{\diff z}{\diff \lambda})=\mbox{constant},\qquad \frac{\diff z}{\diff \lambda} = \pm\sqrt{-(A^{-1})^{\mu\nu}p_\mu p_\nu}.
\end{equation}
We recognize the usual relation between the translation Noether charges and the tangent vector, which is null:
\begin{equation} \label{geodesic AB}
    p_\mu= g_{\mu\bnu}\frac{\diff X^\bnu}{\diff \lambda}=\mbox{constant},\qquad g_{\bmu\bnu}\frac{\diff X^\bmu}{\diff \lambda}\frac{\diff X^\bnu}{\diff \lambda} =0.
\end{equation}
The two formulas in (\ref{geodesic AB}) would normally be the starting point for solving the geodesic equation in the translation-invariant metric (\ref{bulk metric ansatz}); they can be verified to imply (\ref{bulk metric reconstructed}).

Thus, the condition (\ref{bulk metric assumption}) is sufficiently strong to recover (null) Riemannian geometry in the bulk!

\subsubsection{Proof in the general case}
Let us now consider the general case, without isometries. We assume that we are given a family of codimension-one boundary hyperboloids $x^\mu(p; X)$, indexed by $X$. With no loss of generality (at generic points), we can assume that the surface is parametrized by its normal $p_\mu$, which defines $x^\mu(p; X)$ as a homogeneous function of $p$. The main physical assumption is that, for any $p$, we can find a small displacement that satisfies the tangent condition (\ref{inclusion saturation}), and that the collection of these displacements satisfies a \emph{quadratic} equation:
\begin{equation} \label{assumption general}
    \exists \, g_{\bmu\bnu}(X), \quad \mbox{such that} \quad \forall \, p, \quad \frac{\partial x^\mu(p;X)}{\partial X^\bmu}\fdiff X^\bmu=0 \quad \mbox{implies} \quad g_{\bmu\bnu}(X)\fdiff X^\bmu \fdiff X^\bnu=0.
\end{equation}
We additionally make the technical assumptions that $g_{\bmu\bnu}(X)$ is non-degenerate and that all $d\times(d+1)$ matrices we encounter have the maximal possible rank, $d$. In the preceding formula, this means that $\fdiff X^\bmu(p; X)$ is unique up to rescaling.

We now show that any such family of boundary curves originates from the null geodesics of some bulk metric, namely $g$.

\paragraph{Existence of a Hamilton-Jacobi function} The first step is to show that the following defines a sensible Hamilton-Jacobi function:
\begin{equation} \label{def S invert}
    S(p;X)\equiv p_\mu x^\mu(p;X). 
\end{equation}
Indeed, differentiating with respect to $p$, we find
\begin{align}
    \frac{\partial S(p;X)}{\partial p_\mu} &= x^\mu(p;X) + p_\nu \frac{\partial x^\nu(p;X)}{\partial p_\mu} \\
    &= x^\mu(p;X),
\end{align}
as required by (\ref{defs}). The second term on the first line vanishes because $p$ is assumed to be normal. This shows that there is nothing exceptional about surfaces that can be written as the gradient of an action $S$: one just has to be able to parametrize the surface by its normal (in some coordinates).

\paragraph{Derivation of Hamilton-Jacobi equation}
We now want to prove that $S$ defined by (\ref{def S invert}) satisfies the Hamilton-Jacobi equation. The essential step is the reverse of the proof of (\ref{dX prop P}). Namely, we dot (\ref{assumption general}) into $p_\mu$ and differentiate with respect to $p_\nu$:
\begin{equation}
    0=\frac{\partial}{\partial p_\nu}\qty(
    \frac{\partial S(p;X)}{\partial X^\bmu} \fdiff X^\bmu(p;X))=
    \frac{\partial S(p;X)}{\partial X^\bmu}
    \frac{\partial \fdiff X^\bmu(p;X)}{\partial p_\nu} + \cancel{\frac{\partial^2S(p;X)}{\partial p_\nu \partial X^\bmu} \fdiff X^\bmu(p;X)}.
\end{equation}
The second term is equal to $\frac{\partial x^\nu}{\partial X^\bmu} \fdiff X^\bmu$ and vanishes for the displacements considered in (\ref{assumption general}). Now, taking a derivative with respect to $p_\nu$ of the null condition in (\ref{assumption general}) we get that
\begin{equation}    
    g_{\bmu\bnu}(X)\fdiff X^\bnu(p;X) \frac{\partial\fdiff X^\bmu(p;X)}{\partial p_\nu}=0.
\end{equation}
Assuming that the matrix on the right, $\frac{\partial\fdiff X^\bmu}{\partial p_\nu}$, has the maximal rank, $d$, comparing the two equations tell us that
\begin{equation}
    \frac{\partial S(p;X)}{\partial X^\bmu}\propto g_{\bmu\bnu}(X)\fdiff X^\bnu(p;X),
\end{equation}
so that the null condition can be written equivalently as
\begin{equation}
    g^{\bmu\bnu}(X)\frac{\partial S(p;X)}{\partial X^\bmu}\frac{\partial S(p;X)}{\partial X^\bnu}=0,
\end{equation}
which is the Hamilton-Jacobi equation~(\ref{HJ}) as desired. We conclude that the condition (\ref{assumption general}), namely that bulk displacements that lead to tangent boundary changes satisfy a quadratic equation, is equivalent to the Hamilton-Jacobi equation. As shown in section~\ref{s:hj-wkb}, this implies that the boundary curves themselves can be constructed by integrating the null geodesic equation.

It is interesting to ask what principle in the boundary theory might explain the quadratic property in (\ref{assumption general}). Here we simply note that this property is automatic by symmetry for bulk points sufficiently close to the AdS boundary (see \eqref{hyperbola}). Thus, the essential question is to understand why bulk physics is the same around every point: the equivalence principle.

\section{Conclusion}\label{s:discussion}
In this paper we solved an exercise in geometrical optics: we studied null geodesics in a spacetime with an (asymptotic) boundary. Geodesics can be characterized by either their phase space at the boundary $(x,p)$ or in the bulk $(X, P)$; in the Hamilton-Jacobi formalism, one gets to fix either the position or the momentum at each endpoint. We found that the most useful way to label geodesics is to fix a bulk point $X$ and the momentum $p$ with which the geodesic reaches the boundary. The resulting Hamilton-Jacobi function $S(p; X)$ encodes all relevant optical properties of the bulk as seen from the boundary.

Our physical motivation for this problem is to study boundary correlation functions integrated against directional wavepackets, which will be pursued in \cite{Caron-Huot:2025tap}. A key fact about wavepacket scattering is that the boundary data vastly over-determines the position of a putative bulk point where particles can meet and experience a local scattering event. Random wavepackets miss. This is so even for the simplest case of four-point correlators, as depicted in figure~\ref{fig:4pt-lightraysA}. The \emph{boundary hyperboloids} $H^\pm(X)$ (related to the ``light-cone cuts'' of \cite{Engelhardt:2016wgb, Engelhardt:2016vdk}) characterize when singular features can occur: when multiple operators approach the future or past lightcone of a common bulk point. In principle, hyperboloids can be measured locally: one only needs access to a small open neighborhood of a single point in $H^-$ and of a single point in $H^+$.

In this setup, the measured hyperboloids are not mathematical singularities but are more like bumps that can get blurry if the theory is not of holographic type. Necessary conditions for the correlators of a CFT to display sharp features, discussed in \cite{Caron-Huot:2022lff}, include the requirements of \cite{Heemskerk:2009pn}: the CFT must have a large central charge and a large higher-spin gap.  In this paper, we treated the ideal case of sharp lightcones which appear in this large-parameter limit.

Of course, even if a theory's dynamics are auspicious, there can always exist bulk points that are hard to ``see'' from the boundary. A pleasing finding in this paper is that many interesting locations, including points close to (but outside) the horizon of small or large black holes, can be observed from simple boundary correlation functions without particular effort.

Boundary hyperboloids enjoy interesting mathematical properties. They are simply parametrized by the gradient of the Hamilton-Jacobi function $S(p; X)$ mentioned above, and $p$ coincides with the normal vector at each point of the surface. The extrinsic curvature of the surface provides a natural notion of depth: the parallax, i.e., the relation between the arrival position and angle of signals from $X$. Finally, bulk causality has a striking signature: boundary hyperboloids corresponding to timelike separated bulk points must be \emph{included} in each other.

As was also observed in \cite{Engelhardt:2016crc} and explicated in section~\ref{s:inclusion}, the inclusion property gives a simple experimental method to measure the conformal metric near a bulk point $X$. One may ask if the overall scale factor can also be measured using the same correlators. Indeed, this can be achieved in many ways using any physical ruler provided by the bulk theory, such as a dimensionful coupling entering a bulk scattering amplitude or the mass of a produced resonance \cite{Caron-Huot:2022lff, Caron-Huot:2025tap}.

A remarkable feature is that the inclusion property can be turned around. Given a family of boundary hyperboloids associated with bulk points $X+\fdiff X$, null bulk displacements $\fdiff X$ can be identified by looking for tangent points on the hyperboloids. As shown under some genericity assumptions in \S\ref{ssec:reverse}, this gives a physical interpretation to the bulk Hamilton-Jacobi equation: it is simply the statement that the null vectors $\fdiff X$ satisfy a common quadratic equation~(which defines the conformal metric at the point $X$). In this way, the geodesic equation in the bulk is derived from the geometry of boundary curves.  This is somewhat reminiscent of the emergence of bulk geometry from causal diamonds \cite{Czech:2016xec,deBoer:2016pqk} or entanglement (see for example \cite{VanRaamsdonk:2010pw}) but it seems distinct. It is an outstanding question to explain how the inclusion property emerges from strong interactions in the boundary CFT.

While Einstein's equations played no role in this paper, it is natural to ask how bulk Raychaudhuri equations and focusing theorems influence the boundary hyperboloids---and eventually, if these can be derived from general CFT properties.

Previously, geometric optics has been quite instrumental in studying physics in black hole backgrounds. On a similar note, properties of null geodesics have proven quite useful in understanding chaotic behavior in different spacetimes \cite{Berenstein:2023vtd}. Our present paper lays the essential groundwork for future investigations within these avenues, which includes analyzing bulk features in more complex backgrounds such as in non-static and time-dependent spacetimes.

\section*{Acknowledgments}
We thank Nima Afkhami-Jeddi, Miguel Correia, and Viraj Meruliya for useful discussions. The work of S.C.H. and J.C. is supported by the National Science and Engineering Council of Canada (NSERC) and the Canada Research Chair program, reference number CRC-2022-00421. K.N. is supported in part by the Natural Sciences and Engineering Research Council of Canada (NSERC), funding reference number SAPIN/00047.

\bibliographystyle{JHEP}
\bibliography{refs}

\end{document}